\documentclass[apjl,iop]{emulateapj}

\usepackage{graphicx}

\newcommand{\sm}[1]{\mbox{{\scriptsize #1}}}

\newcommand{\bef}{\begin{figure}}
\newcommand{\eef}{\end{figure}}

\def\eps@scaling{0.95}

\def\showone#1{
  \centering
  \leavevmode
  \epsfxsize=\eps@scaling\linewidth
  \epsfbox{#1.eps}
}

\def\showtwover#1#2{
  \centering
  \leavevmode
  \epsfxsize=\eps@scaling\linewidth
  \epsfbox{#1.eps} \hfil
  \epsfxsize=\eps@scaling\linewidth
  \epsfbox{#2.eps}
}

\def\showthreeover#1#2#3{
  \centering
  \leavevmode
  \epsfxsize=\eps@scaling\linewidth
  \epsfbox{#1.eps} \hfil
  \epsfxsize=\eps@scaling\linewidth
  \epsfbox{#2.eps} \hfil
  \epsfxsize=\eps@scaling\linewidth
  \epsfbox{#3.eps}
}

\def\showfourover#1#2#3#4{
  \centering
  \leavevmode
  \epsfxsize=\eps@scaling\linewidth
  \epsfbox{#1.eps} \hfil
  \epsfxsize=\eps@scaling\linewidth
  \epsfbox{#2.eps} \hfil
  \epsfxsize=\eps@scaling\linewidth
  \epsfbox{#3.eps} \hfil
  \epsfxsize=\eps@scaling\linewidth
  \epsfbox{#4.eps}
}

\def\epstwo@scaling{0.46}

\def\showtwo#1#2{
  \centering
  \leavevmode
  \epsfxsize=\epstwo@scaling\linewidth
  \epsfbox{#1.eps} 
  \epsfxsize=\epstwo@scaling\linewidth
  \epsfbox{#2.eps}
}

\def\epsthree@scaling{0.28}
\def\showthree#1#2#3{
  \centering
  \leavevmode
  \epsfysize=\epsthree@scaling\textwidth 
  \epsfbox{#1.eps} 
  \epsfysize=\epsthree@scaling\textwidth 
  \epsfbox{#2.eps}
  \epsfysize=\epsthree@scaling\textwidth 
  \epsfbox{#3.eps}
}

\def\epstwo@scaling{0.44}

\def\showfour#1#2#3#4{
  \centering
  \leavevmode
  \epsfxsize=\epstwo@scaling\linewidth
  \epsfbox{#1.eps} \hfil
  \epsfxsize=\epstwo@scaling\linewidth
  \epsfbox{#2.eps} \hfil
  \epsfxsize=\epstwo@scaling\linewidth
  \epsfbox{#3.eps} \hfil
  \epsfxsize=\epstwo@scaling\linewidth
  \epsfbox{#4.eps}
}

\def\showsix#1#2#3#4#5#6{
  \centering
  \leavevmode
  \epsfxsize=\epstwo@scaling\linewidth
  \epsfbox{#1.eps} \hfil
  \epsfxsize=\epstwo@scaling\linewidth
  \epsfbox{#2.eps} \hfil
  \epsfxsize=\epstwo@scaling\linewidth
  \epsfbox{#3.eps} \hfil
  \epsfxsize=\epstwo@scaling\linewidth
  \epsfbox{#4.eps} \hfil
  \epsfxsize=\epstwo@scaling\linewidth
  \epsfbox{#5.eps} \hfil
  \epsfxsize=\epstwo@scaling\linewidth
  \epsfbox{#6.eps}
}

\newcommand{\befone}{
  \begin{figure*}
  \centering
  \begin{minipage}{\textwidth}
  }
\newcommand{\eefone}{\end{minipage}\end{figure*}}

\shorttitle{Black hole formation in primordial galaxies.}
\shortauthors{Schleicher et al.}

\begin{document}


\title{Black hole formation in primordial galaxies: chemical and radiative conditions}


\author{Dominik R. G. Schleicher\altaffilmark{1,2}, Marco Spaans\altaffilmark{3}, Simon C. O. Glover\altaffilmark{4}}
\email{dschleic@eso.org}
\altaffiltext{1}{ESO Garching, Karl-Schwarzschild-Str. 2, 85748 Garching bei M\"unchen, Germany}
\altaffiltext{2}{Leiden Observatory, Leiden University, P.O.Box 9513, NL-2300 RA Leiden, the Netherlands}
\altaffiltext{3}{Kapteyn Astronomical Institute, University of Groningen, P.O. Box 800, 9700 AV, Groningen, the Netherlands}
\altaffiltext{4}{Institut f\"ur Theoretische Astrophysik, Albert-Ueberle-Str. 2, D-69120 Heidelberg, Germany}

\begin{abstract}
In massive primordial galaxies, the gas may directly collapse and form a single central massive 
object if cooling is suppressed. H$_2$ line cooling can be suppressed in the presence
of a strong soft-ultraviolet radiation field, but the role played by other cooling mechanisms
is less clear. In optically thin gas,  Lyman-$\alpha$ cooling can be very effective, maintaining the gas 
temperature below $10^{4} \: {\rm K}$ over many orders of magnitude in density. However, the large
neutral hydrogen column densities present in primordial galaxies render them highly optically thick
to Lyman-$\alpha$ photons. In this paper, we examine in detail the effects of the trapping of these
Lyman-$\alpha$ photons on the thermal and chemical evolution of the gas. We show that despite 
the high optical depth in the Lyman series lines, cooling is not strongly suppressed, and proceeds
via other atomic hydrogen transitions.
 At densities larger than $\sim10^9$~cm$^{-3}$, collisional dissociation of molecular hydrogen becomes the dominant cooling process and decreases the gas temperature to about $5000$~K. The gas temperature evolves with density as $T \propto \rho^{\gamma_{\rm eff} 
- 1}$, with $\gamma_{\rm eff} = 0.97-0.98$. The evolution is thus very close to isothermal, and so fragmentation is possible, but unlikely to occur during the initial collapse. However, after the 
formation of a massive central object, we expect that later-infalling, higher angular momentum material
will form an accretion disk that may be unstable to fragmentation{, which may give rise to star formation with a top-heavy IMF}.
\end{abstract}


\section{Introduction}
Supermassive black holes with masses $M > 10^{8}$--$10^{9} \: {\rm M_{\odot}}$ 
are known to have existed at very early times in 
the history of the universe \citep{Jiang09}, prompting one to ask how black holes of such a size 
could have formed so quickly. {They may have} started life as 
stellar-mass objects, the compact remnants of the first generation of stars, and they {may} have 
grown to their present size by accretion. However, forming supermassive black holes in this
way presents a number of difficulties. The first stars are thought to have had masses of order
$100 \: {\rm M_{\odot}}$ or more, making them strong sources of ionizing radiation 
\citep{Abel02, Bromm04, Glover05,Yoshida08}. The dark matter halos in which these stars 
formed were small, with masses of only $10^{6} \: {\rm M_{\odot}}$, and models of the effects 
of the ionizing radiation produced by a primordial star within one of these small halos show 
that it readily photo-evaporates the bulk of the gas from the halo, leaving little gas available to 
be accreted by the black hole \citep{Johnson07, Milosavljevic08,Alvarez09}. {Although
accretion is likely more efficient in the $5-6$~$\sigma$ peaks that harbor the observed supermassive 
black holes, it is still constrained by the Eddington limit. As discussed by \citet{Shapiro05}, seed masses
of at least $10^5\ M_\odot$ are required for an MHD disk or a standard thin disk.  Similarly, \citet{SchleicherSpaans10} found that high Eddington ratios and unusually low mass-energy conversion efficiencies  are needed for stellar progenitors.
}
%




Previous studies have therefore examined the possibility of forming these large seed black holes
by direct gravitational collapse \citep{Eisenstein95, Koushiappas04, Begelman06, Spaans06}. 
Gas falling into a halo with a mass $M > 5 \times 10^{7} [(1 + z) / 10]^{-3/2} \: {\rm M_{\odot}}$
(corresponding to a virial temperature $T_{\rm vir} > 10^{4} \: {\rm K}$) will be shock-heated 
up to $T_{\rm vir}$, and so will become hot enough to cool 
via the electronic emission lines of atomic hydrogen. The fate of the gas in one of these so-called
`atomic cooling' halos will then depend on its subsequent thermal evolution. If the gas can cool efficiently as it collapses, significantly lowering both its temperature and its Jeans mass, then 
it is very difficult to prevent it from fragmenting, and hence forming stars 
rather than a massive black hole \citep{Bromm03, Omukai08, Clark09}. On the other hand, if cooling remains inefficient {and the} effective equation of state for the gas is stiffer than isothermal, then 
fragmentation is suppressed \citep{LiMacLowKlessen05, Lodato06} and the formation of a massive
black hole is a plausible outcome. In gas that has been enriched with heavy elements and dust from
the first generation of stars, it appears that cooling and fragmentation cannot be avoided 
\citep{Omukai08}. On the other hand, in primordial gas the only effective low-temperature coolant
is molecular hydrogen (H$_{2}$), and if enough of this can be destroyed by photodissociation, then 
cooling can be suppressed \citep{Begelman06,Dijkstra08,Regan09,Shang09}. {The soft-UV background needs to be strong enough to suppress H$_2$ formation until densities of $\sim10^5$~cm$^{-3}$ are reached, following which collisional dissociation will suppress the H$_2$ abundance for gas temperatures greater than a few $1000$~K. The required background field depends on the chemical model, the spectral shape and the potential presence of additional electrons created in shock fronts. It is of the order $J_{21}=10^3-10^5$ \citep{Omukai01, Bromm03, Shang09}.}
These values are much larger than estimates of the mean 
strength of the background, which range from $J_{21} \sim 0.1$ \citep{Johnson08a} to 
$J_{21} \sim 40$ \citep{Dijkstra08}. However, recent work has shown that the radiation background 
is highly inhomogeneous, owing to the strong clustering of the first generation of star-forming 
galaxies \citep{Dijkstra08,Ahn09}. In regions close to these galaxies, the ultraviolet background 
can be very large, and \citet{Dijkstra08} have recently demonstrated that enough massive primordial galaxies {with a local value of $J_{21}>10^3$ may exist.} In addition, early reionization in local patches of the universe may suppress star formation in some primordial halos in a similar fashion \citep{Johnson09}. They may thus remain 
pristine, but grow in mass until they can collapse.

In the absence of H$_{2}$ cooling, emission from the Lyman series lines of atomic hydrogen,
especially Lyman-$\alpha$, plays a central role in regulating the temperature of the collapsing 
gas. Hydrodynamical models of the collapsing gas typically assume that 
the Lyman series lines remain effectively optically thin, allowing the optically thin form of the 
cooling rate to be used. These models find that the temperature of the collapsing gas decreases
with increasing density, albeit very slowly, typically reaching $T \sim 6000$--7000~K at densities
of order $n \sim 10^{9} \: {\rm cm^{-3}}$. However, the large hydrogen column densities present 
in these protogalaxies produce very large optical depths in the Lyman series lines, and so the assumption that the optically thin form 
of the cooling rate can safely be used is highly questionable. \citet{Spaans06} examined the 
effects of  Lyman-$\alpha$ photon trapping {within an analytic framework} for the evolution of the gas and showed that in the absence of additional cooling 
mechanisms, it gives rise to a temperature evolution described by 
\begin{equation}
\gamma_{\rm eff} =1+\frac{d\log T}{d\log \rho}\sim1-\frac{0.5-7/18\cdot Bn^{7/18}}{\log C n^{1/2}+Bn^{7/18}},
\end{equation}
where $\rho$ is the mass density, $T$ the temperature, $n$ the number density, $B\sim0.5-0.1$~cm$^{7/6}$ and $C\sim10^{-36}M_h/(10^7\ M_\odot)$~cm$^{3/2}$, with $M_h$ the halo mass. For halo
masses in the range $10^{7}$--$10^{9} \: {\rm M_{\odot}}$, 
this yields $\gamma_{\rm eff} -1\sim0.01-0.5$, indicating an effective equation of
state that is stiffer than isothermal, and hence a temperature that {\em increases} with increasing density.
This result suggests that not only is the neglect of the effects of photon trapping a poor approximation,
but it may also lead to qualitatively incorrect results, {as it is crucial for fragmentation whether the effective equation of state is harder than isothermal.}


In this paper, we re-examine the issue of Lyman-$\alpha$ photon trapping, using a considerably
more detailed treatment than in \citet{Spaans06}. In particular, we address the issue of whether 
there are other cooling mechanisms that can compensate for Lyman-$\alpha$ if the latter is strongly
suppressed. 

\section{Method}\label{methods}
To follow the thermal and chemical evolution of the collapsing gas, we make use of a one-zone
treatment in which the form of the density evolution is prescribed in advance. One-zone models
are widely used for studying the chemistry and thermodynamics of primordial or low metallicity
gas \citep[e.g.][]{Omukai01,Omukai05,Cazaux09,Glover09,Schleicher09b}, as they enable one to 
model the chemistry in great detail and to include a large number of different cooling processes, 
without the computational efficiency concerns inherent to a three-dimensional treatment.

We extend the one-zone model developed by \citet{Glover09} and \citet{Schleicher09b} to include effects that become important at higher temperatures and in the presence of Lyman-$\alpha$ trapping. Near $T\sim10^4$~K, H$^-$ formation cooling may contribute to the overall cooling rate \citep{Omukai01}. We assume that a typical electron undergoing radiative attachment to form H$^{-}$
has an energy of order $k_{B} T$, where is $k_B$ Boltzmann's constant, and hence write the 
H$^{-}$ formation cooling rate as
\begin{equation}
\Lambda_{\rm H^-}=k_{\rm H^-} n_{\rm H} n_{\rm e} k_B T,
\end{equation}
where $k_{\rm H^-}$ is the H$^-$ formation rate coefficient, $n_{\rm H}$ is the atomic hydrogen number  density, and $n_{\rm e}$ is the electron number density. To model the effects of Lyman-$\alpha$ trapping, we include different level populations as separate species in the code. In our model, we consider energy levels up to $n=5$. For the first excited state, we distinguish between the $2$s and the $2$p states, as only decays from the $2$p state to the ground state will produce Lyman-$\alpha$ photons; two-photon
decays from the $2$s state will produce continuum photons that will not be trapped. The other states 
are considered as averages over the angular momentum quantum numbers. Radiative decay rates 
$A_{ij}$, collisional excitation and de-excitation rates $C_{ij}$, collisional ionization and three-body recombination rates, and collisional transition rates between the $2$s and $2$p state are all
adopted from \citet{Omukai01}. We define the transition rate $R_{ij}$ from level $i$ to level $j$ as
\begin{eqnarray}
R_{ij}&=&A_{ij}\beta_{\sm{esc},ij}(1+Q_{ij})+C_{ij},\quad  i>j\\
R_{ij}&=&\frac{g_j}{g_i}A_{ji}\beta_{\sm{esc},ji}Q_{ji}+C_{ij}, \quad i<j,
\end{eqnarray}
where $\beta_{\sm{esc},ij}$ is the escape probability for the $i \rightarrow j$ transition, $g_i=2i^2$ denotes the statistical weight of level $i$ and $Q_{ij}=c^2 J_{\sm{cont},ij}/(2h\nu_{ij}^3)$, where 
$\nu_{ij}$ is the frequency of the transition and $J_{\sm{cont},ij}$ the average intensity of the background radiation field at this frequency. The spectral shape of the background radiation field
is taken to be that of a $10^5 \: {\rm K}$ \citep{Glover09}, and the normalization is set by
fixing the value of  $J_{21}$. We assume that UV photons more energetic than 13.6~eV are 
absorbed by atomic hydrogen in the intergalactic medium, and we do not consider the effects of
an extragalactic X-ray background.  
For radiative or collisional transitions to the first excited state, we assume that the reaction products will be distributed according to the statistical weights of the 2s and 2p states. However, our results are not sensitive 
to this assumption, as collisional transitions between these two states occur rapidly in the conditions
of interest, and so the ratio between the 2s and 2p level populations is always very close to equilibrium.

For the escape probability $\beta_{\sm{esc},ij}$, we adopt the expression
\begin{equation}
\beta_{\sm{esc},ij}=\frac{1-\mathrm{exp}(-\tau_{ij})}{\tau_{ij}}\mathrm{exp}(-\beta t_{\sm{ph}} / t_{\sm{coll}}),
\end{equation}
where $\tau_{ij}$ is the optical depth at line centre of the $i \rightarrow j$ transition,
$t_{\sm{coll}}$ the collapse time of the gas, and $t_{\sm{ph}}$ is the photon diffusion time, i.e.\ the
time required for a photon to diffuse out of the optically thick gas{, and $\beta$ a geometrical factor. Even small deviations from spherical symmetry lead to escape along a preferred direction and $\beta=3$ \citep{Dijkstra06a, Spaans06}.} In our model, the collapse time is the free-fall time, corrected by a factor that takes into account the effective equation of state \citep[see][]{Omukai05, Schleicher09b}. For most line transitions, $t_{\sm{ph}} \ll t_{\sm{coll}}$, such that the exponential factor is negligible. {For these lines, we follow \citet{Omukai00} and assume that the dominant contribution to the optical depth comes from material within one local Jeans length, and that the density, temperature etc.\ do not vary significantly on this scale. This is justified provided that the collapse is close to isothermal, and that the initial mass of the collapsing gas is comparable to the Jeans mass, as in this case the density and velocity profiles of the gas will come to resemble the Larson-Penston similarity solution \citep{Larson69,Penston69}, which has just these properties.}
For Lyman-$\alpha$ photons and other direct transitions to the ground state, we compute the optical depth $\tau$ and the photon 
diffusion timescale following \citet{Spaans06}, in order to take into account the significant line broadening that occurs during the diffusion process. For the Lyman-$\alpha$ line, this yields a 
timescale $t_{\sm{ph}}=L (a\tau_{21})^{1/3}/c$ with the natural-to-thermal line-width $a$
and an optical depth $\tau_{21}=1.04\times10^{-13}N_{\rm H} T_4^{-0.5}$, with $L$ the distance
to the edge of the halo, and $N_{\rm H}$ the column density of atomic hydrogen. For the Lyman-$\alpha$ line, the natural-to-thermal line-width is taken as $a=4.7\times10^{-4}T_4^{-1/2}$ \citep{Spaans06}, while we correct this value for the increased lifetime in case of other line transitions to the ground state. 
To calculate the column density, we assume that the density profile of the gas scales with radial distance $r$ as $r^{-2.2}$, as indicated by previous numerical simulations \citep[e.g.][]{Wise07a}. The cooling functions for the hydrogen lines are then evaluated based on the escape probability, the level populations and the strength of the background radiation.


\section{Results and conclusions}\label{results}

\begin{figure}[t]
\includegraphics[scale=0.45]{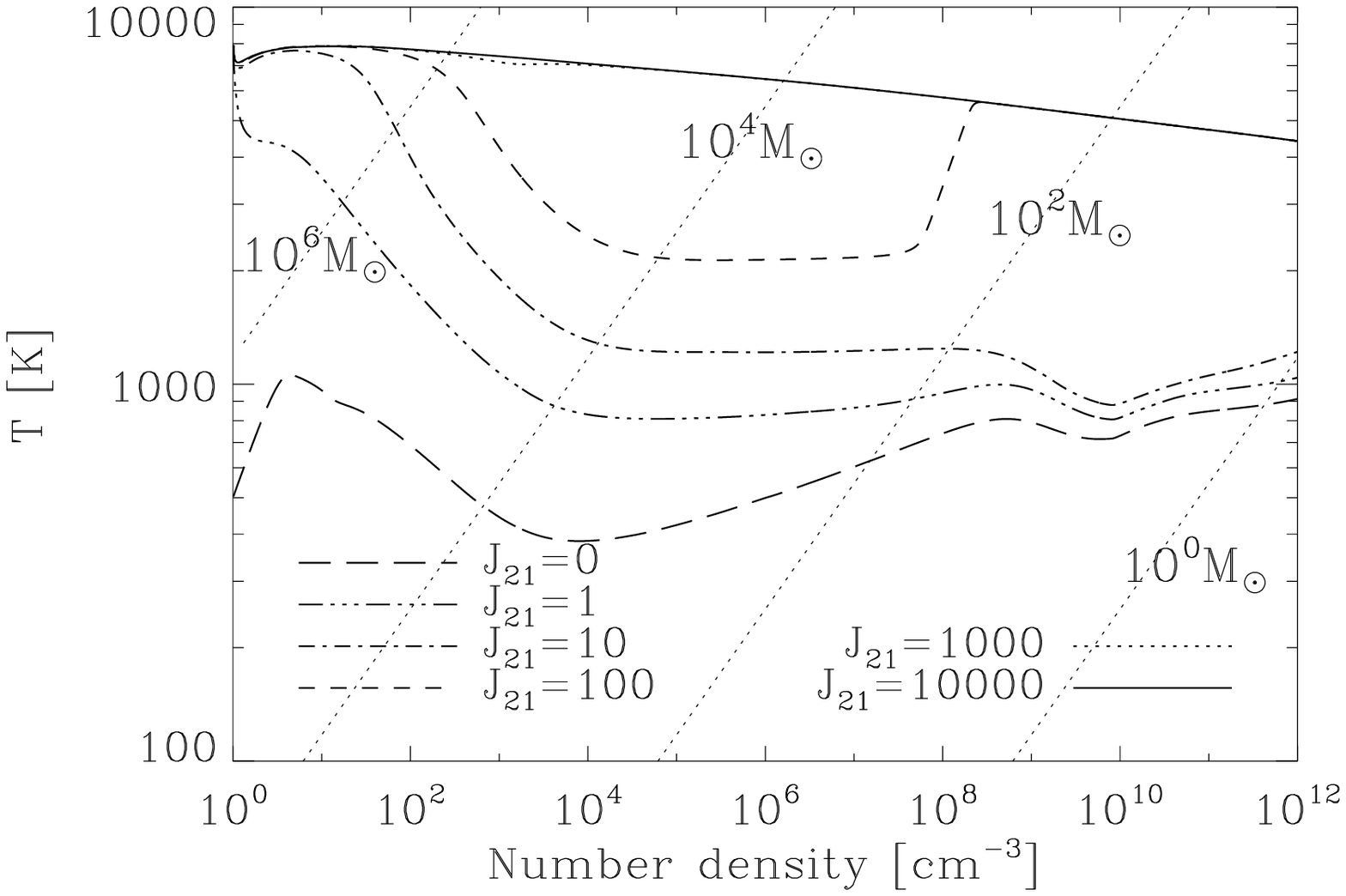}
\caption{Temperature evolution as a function of density for different values of $J_{21}$. The thin dotted lines indicate
lines of constant Jeans mass. 
%
}
\label{fig:temp}
\eef

\begin{figure}[t]
\includegraphics[scale=0.45]{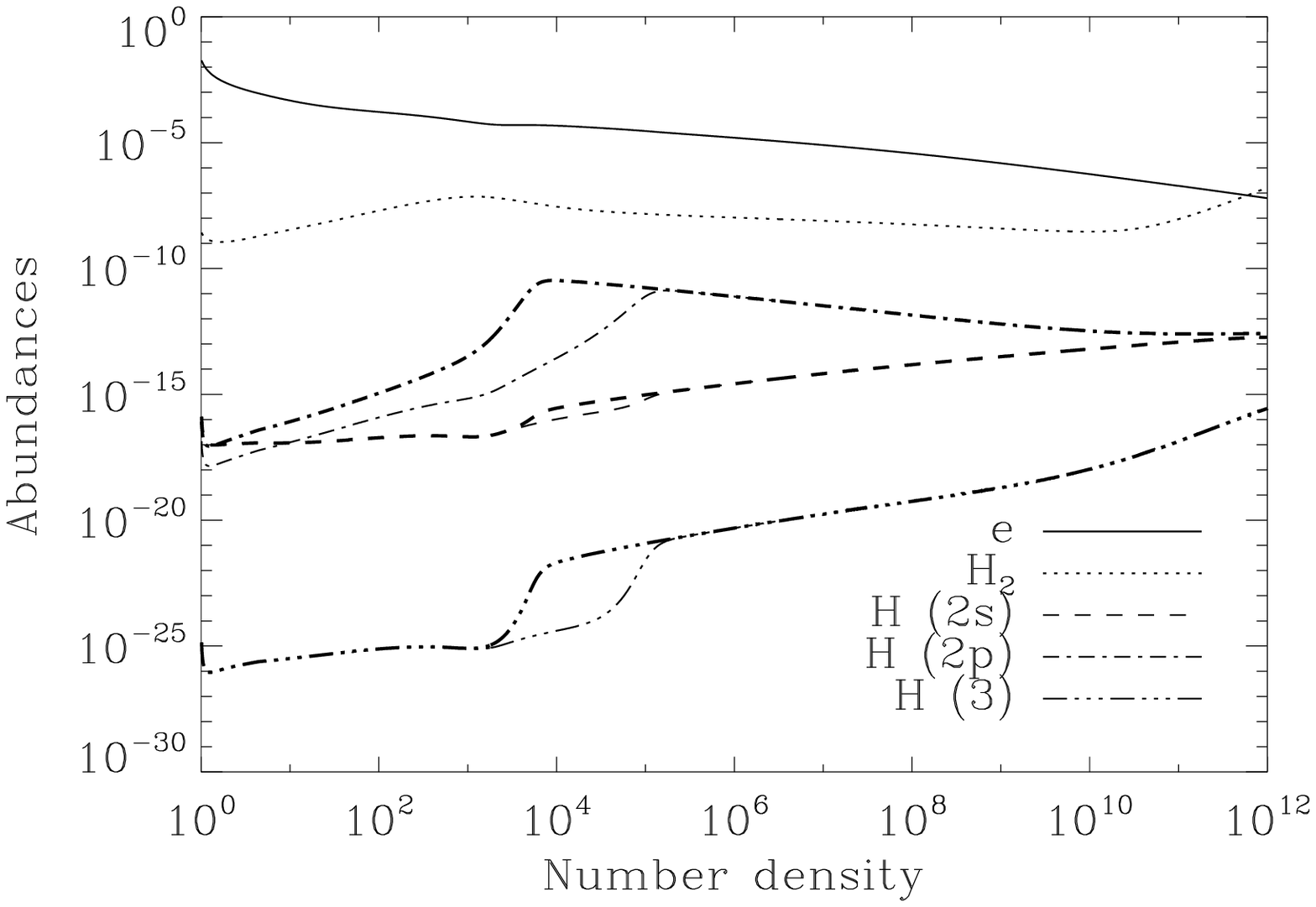}
\caption{Fractional abundances for electrons, H$_2$ and the hydrogen level populations, plotted as
a function of density, for the case where $J_{21}=1000$. The values of the hydrogen level populations
depend on the assumed size of the protogalaxy; values are plotted for virial radii of $500$~pc (thin 
lines) and $3$~kpc (thick lines).}
\label{fig:abun}
\eef

\begin{figure}[t]
\includegraphics[scale=0.45]{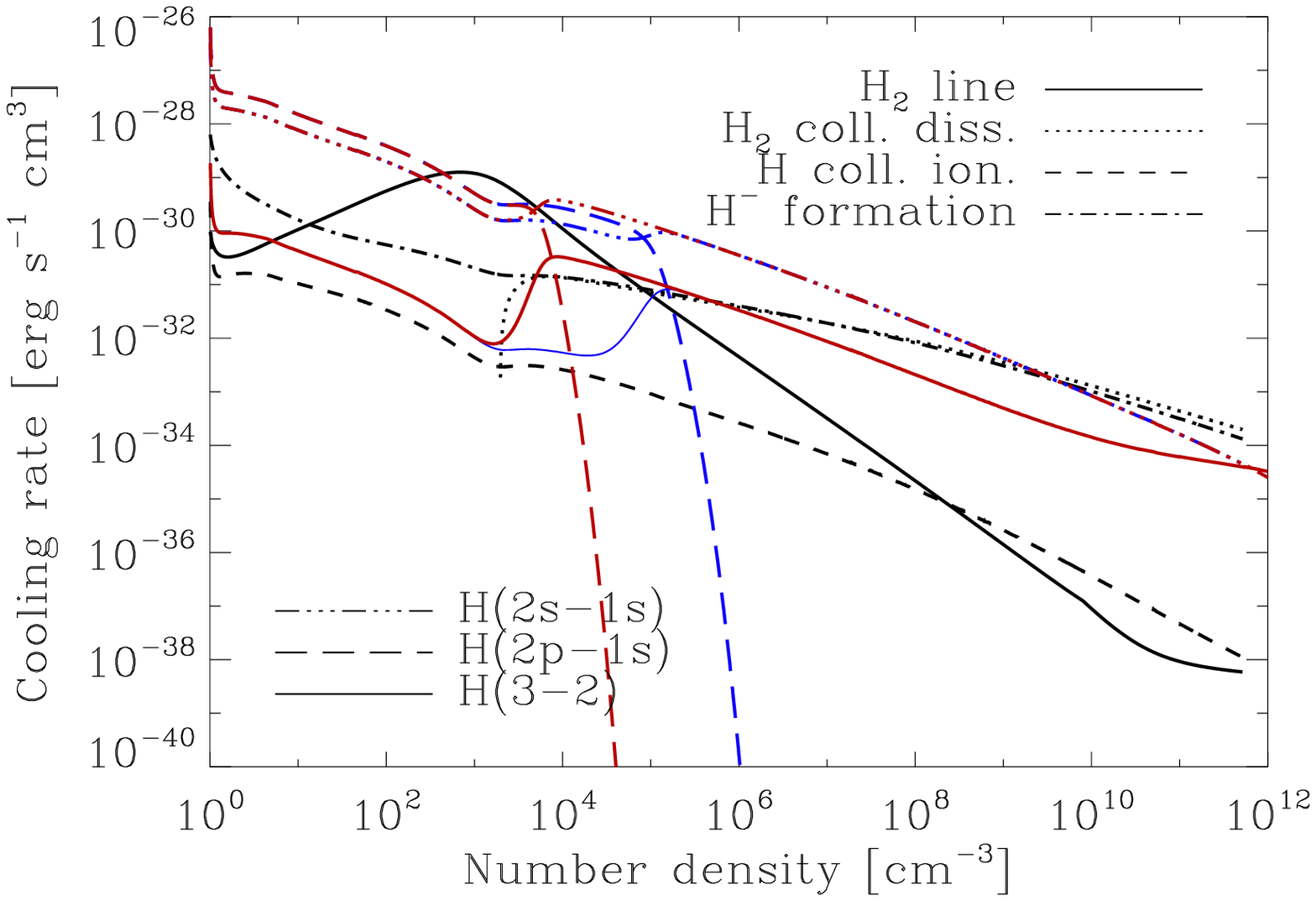}
\caption{The main cooling functions for $J_{21}=1000$ as a function of density. Lines are black if the contribution is not affected by Lyman $\alpha$ trapping. For the hydrogen lines, blue lines correspond to a system with virial radius $0.5$~kpc ($\sim10^7\ M_\odot$), and red lines to a virial radius of $3$~kpc ($\sim2\times10^9\ M_\odot$). The H$_2$ collisional dissociation cooling rate is corrected for the effect of H$_2$ heating.}
\label{fig:cool}
\eef

The temperature evolution that we find with our approach for several different values of $J_{21}$ is plotted as a function of density in Fig.~\ref{fig:temp}. Even for $J_{21}=1$, the temperature remains much higher during the collapse than in the radiation-free case, and never drops significantly below 
$1000$~K. 

As previously noted, much higher values of $J_{21}$ can be obtained locally, in the presence of a luminous neighbour within a distance of $10$~kpc or less \citep{Dijkstra08}. As we increase $J_{21}$,
we find systematically higher temperatures at each density, and for $J_{21} = 100$, we find that the
effects of H$_{2}$ collisional dissociation start to become important. In this case, H$_{2}$ cooling is
marginally effective at densities $n < 10^{8} \: {\rm cm^{-3}}$, but at higher densities, collisional
dissociation destroys the H$_{2}$, causing a sharp rise in the temperature. Such a radiation field 
may be present in a fraction of $10^{-5}-10^{-3}$ of all atomic cooling halos \citep{Dijkstra08}.
For $J_{21}=10^3$ and $J_{21}=10^4$, we find, in common with previous studies, that H$_{2}$
cooling never becomes important, and the gas temperature remains high. Nevertheless, in contrast 
to the predictions of \citet{Spaans06}, we find that the temperature does decline with increasing 
density, although it never drops below $5000$~K over the range of densities examined here. With $\gamma_{\rm eff} =0.97-0.98$,  
we find that the effective equation of state of the gas is just slightly softer than isothermal. {We note, though, that this estimate provides just a lower limit for the value of $\gamma_{\rm eff}$, which is strictly defined as $d\log T/d\log\rho$ at constant entropy. }

In order to understand why the gas is still able to cool, we explore the case with $J_{21}=1000$ in 
more detail in Fig.~\ref{fig:abun} and Fig.~\ref{fig:cool}. In Figure~\ref{fig:abun}, we show the evolution
of the fractional abundances of free electrons and H$_{2}$ molecules, as well as the fractional level
populations of the 2s, 2p and 3 states of atomic hydrogen (denoted henceforth as H(2s), H(2p) 
and H(3)). In Figure~\ref{fig:cool}, we show the contributions that the main cooling processes make
to the total cooling rate. The Lyman-$\alpha$ optical depth depends on the size of the protogalaxy,
and we consider two examples, one with a virial radius of $500$~pc (corresponding to a halo mass of 
$\sim10^7\ M_\odot$), and a second with a virial radius of $3$~kpc (corresponding to a halo mass of 
$2\times10^9\ M_\odot$). In each case, the abundance of the 2p state is initially very small, but increases roughly linearly with increasing density, owing to the increasing importance of 
Lyman-$\alpha$ photon trapping. However, collisional excitation to the H(3) state eventually comes 
to dominate over radiative decay from the 2p state, following which the abundance of the 2p state
remains roughly constant. 

The abundances of the H(2s), H(2p) and H(3) states are clearly larger in the case with the larger virial radius, as Lyman-$\alpha$ photons are trapped more effectively, directly boosting the population of 
the 2p state. The population of the 2s state is then increased due to collisional coupling. The same holds true for the H(3) state. However, the electron and H$_2$ abundances are not affected. Despite these changes in the hydrogen level populations, the total cooling rate does not change significantly.
Cooling due to the 2p--1s line decreases, but this is balanced by additional cooling from the 2s--1s
and 3--2 transitions, due to the increased populations of these states.  At densities higher than $10^8$~cm$^{-3}$, collisional dissociation of H$_2$ becomes the dominant cooling process (even after correcting for H$_2$ formation heating) and cools the gas to about $10^5$~K. H$^-$ formation cooling is found to contribute significantly in a broad range of densities. 

Although we find that the effective equation of state of the gas {may be} softer than isothermal, it is 
nevertheless true that $\gamma_{\rm eff} \sim 1$ throughout the collapse. Moreover,  the temperature evolution of the gas is surprisingly similar to that obtained in previous three-dimensional simulations 
that assume optically thin Lyman-$\alpha$ cooling. {The outcome of the initial collapse may thus be similar to} that found in previous simulations, i.e.\ the formation of a single massive
bound object at the centre of the halo \citep[see e.g.][]{Wise08a, Begelman09, Regan09, Shang09}.
However, it is also highly likely that gas that falls in later, with higher angular momentum, will begin
to build up an accretion disk surrounding this object. The high gas temperatures imply a rapid accretion
flow onto such a disk, which {may} quickly become gravitationally unstable. If the gas in the disk is also
able to cool effectively, which our results suggest will be the case even if Lyman-$\alpha$ cooling is
completely discounted, then we would expect it to fragment and form stars \citep{Levin07}. {The latter may be drawn to the center by dynamical friction and merge with the central black hole \citep{Devecchi09}.} Owing to the high temperatures, we would expect the stars formed in the disk to accrete more rapidly than standard Population III stars, by a factor of ten or more, and hence their final masses may be very large, $M \sim100-500\ M_\odot$.  {However, recent works also indicate the possibility of quasi-stable cold self-gravitating accretion disks that do not fragment, which could feed the central object during the further evolution \citep{Levine08,Begelman09}.}
%


\acknowledgments

The research leading to these results has received funding from the European Community's Seventh Framework Programme (/FP7/2007-2013/) under grant agreement No 229517. DRGS and SCOG thank for subsidies from the Landesstiftung Baden-W{\"u}rttemberg via their program International Collaboration II. {We thank the referee Nick Gnedin for helpful comments that improved our manuscript.}



\clearpage




\clearpage


\begin{thebibliography}{40}
\expandafter\ifx\csname natexlab\endcsname\relax\def\natexlab#1{#1}\fi

\bibitem[{{Abel} {et~al.}(2002){Abel}, {Bryan}, \& {Norman}}]{Abel02}
{Abel}, T., {Bryan}, G.~L., \& {Norman}, M.~L. 2002, Science, 295, 93

\bibitem[{{Ahn} {et~al.}(2009){Ahn}, {Shapiro}, {Iliev}, {Mellema}, \&
  {Pen}}]{Ahn09}
{Ahn}, K., {Shapiro}, P.~R., {Iliev}, I.~T., {Mellema}, G., \& {Pen}, U. 2009,
  \apj, 695, 1430

\bibitem[{{Alvarez} {et~al.}(2009){Alvarez}, {Wise}, \& {Abel}}]{Alvarez09}
{Alvarez}, M.~A., {Wise}, J.~H., \& {Abel}, T. 2009, \apjl, 701, L133

\bibitem[{{Begelman} \& {Shlosman}(2009)}]{Begelman09}
{Begelman}, M.~C., \& {Shlosman}, I. 2009, \apjl, 702, L5

\bibitem[{{Begelman} {et~al.}(2006){Begelman}, {Volonteri}, \&
  {Rees}}]{Begelman06}
{Begelman}, M.~C., {Volonteri}, M., \& {Rees}, M.~J. 2006, \mnras, 370, 289

\bibitem[{{Bromm} \& {Larson}(2004)}]{Bromm04}
{Bromm}, V., \& {Larson}, R.~B. 2004, \araa, 42, 79

\bibitem[{{Bromm} \& {Loeb}(2003)}]{Bromm03}
{Bromm}, V., \& {Loeb}, A. 2003, \apj, 596, 34

\bibitem[{{Cazaux} \& {Spaans}(2009)}]{Cazaux09}
{Cazaux}, S., \& {Spaans}, M. 2009, \aap, 496, 365

\bibitem[{{Clark} {et~al.}(2009){Clark}, {Glover}, {Bonnell}, \&
  {Klessen}}]{Clark09}
{Clark}, P.~C., {Glover}, S.~C.~O., {Bonnell}, I.~A., \& {Klessen}, R.~S. 2009,
  arXiv:0904.3302

\bibitem[{{Devecchi} \& {Volonteri}(2009)}]{Devecchi09}
{Devecchi}, B., \& {Volonteri}, M. 2009, \apj, 694, 302

\bibitem[{{Dijkstra} {et~al.}(2008){Dijkstra}, {Haiman}, {Mesinger}, \&
  {Wyithe}}]{Dijkstra08}
{Dijkstra}, M., {Haiman}, Z., {Mesinger}, A., \& {Wyithe}, S. 2008, ArXiv
  e-prints

\bibitem[{{Dijkstra} {et~al.}(2006){Dijkstra}, {Haiman}, \&
  {Spaans}}]{Dijkstra06a}
{Dijkstra}, M., {Haiman}, Z., \& {Spaans}, M. 2006, \apj, 649, 14

\bibitem[{{Eisenstein} \& {Loeb}(1995)}]{Eisenstein95}
{Eisenstein}, D.~J., \& {Loeb}, A. 1995, \apj, 443, 11

\bibitem[{{Glover}(2005)}]{Glover05}
{Glover}, S. 2005, Space Science Reviews, 117, 445

\bibitem[{{Glover} \& {Savin}(2009)}]{Glover09}
{Glover}, S.~C.~O., \& {Savin}, D.~W. 2009, \mnras, 393, 911

\bibitem[{{Jiang} {et~al.}(2009){Jiang}, {Fan}, {Bian}, {Annis}, {Chiu},
  {Jester}, {Lin}, {Lupton}, {Richards}, {Strauss}, {Malanushenko},
  {Malanushenko}, \& {Schneider}}]{Jiang09}
{Jiang}, L., {Fan}, X., {Bian}, F., {Annis}, J., {Chiu}, K., {Jester}, S.,
  {Lin}, H., {Lupton}, R.~H., {Richards}, G.~T., {Strauss}, M.~A.,
  {Malanushenko}, V., {Malanushenko}, E., \& {Schneider}, D.~P. 2009, \aj, 138,
  305

\bibitem[{{Johnson}(2009)}]{Johnson09}
{Johnson}, J.~L. 2009, ArXiv e-prints: 0911.1294

\bibitem[{{Johnson} \& {Bromm}(2007)}]{Johnson07}
{Johnson}, J.~L., \& {Bromm}, V. 2007, \mnras, 374, 1557

\bibitem[{{Johnson} {et~al.}(2008){Johnson}, {Greif}, \& {Bromm}}]{Johnson08a}
{Johnson}, J.~L., {Greif}, T.~H., \& {Bromm}, V. 2008, \mnras, 388, 26

\bibitem[{{Koushiappas} {et~al.}(2004){Koushiappas}, {Bullock}, \&
  {Dekel}}]{Koushiappas04}
{Koushiappas}, S.~M., {Bullock}, J.~S., \& {Dekel}, A. 2004, \mnras, 354, 292

\bibitem[{{Larson}(1969)}]{Larson69}
{Larson}, R.~B. 1969, \mnras, 145, 271

\bibitem[{{Levin}(2007)}]{Levin07}
{Levin}, Y. 2007, \mnras, 374, 515

\bibitem[{{Levine} {et~al.}(2008){Levine}, {Gnedin}, {Hamilton}, \&
  {Kravtsov}}]{Levine08}
{Levine}, R., {Gnedin}, N.~Y., {Hamilton}, A.~J.~S., \& {Kravtsov}, A.~V. 2008,
  \apj, 678, 154

\bibitem[{{Li} {et~al.}(2005){Li}, {Mac Low}, \& {Klessen}}]{LiMacLowKlessen05}
{Li}, Y., {Mac Low}, M.-M., \& {Klessen}, R.~S. 2005, \apj, 626, 823

\bibitem[{{Lodato} \& {Natarajan}(2006)}]{Lodato06}
{Lodato}, G., \& {Natarajan}, P. 2006, \mnras, 371, 1813

\bibitem[{{Milosavljevic} {et~al.}(2008){Milosavljevic}, {Bromm}, {Couch}, \&
  {Oh}}]{Milosavljevic08}
{Milosavljevic}, M., {Bromm}, V., {Couch}, S.~M., \& {Oh}, S.~P. 2008, ArXiv
  0809.2404

\bibitem[{{Omukai}(2000)}]{Omukai00}
{Omukai}, K. 2000, \apj, 534, 809

\bibitem[{{Omukai}(2001)}]{Omukai01}
---. 2001, \apj, 546, 635

\bibitem[{{Omukai} {et~al.}(2008){Omukai}, {Schneider}, \& {Haiman}}]{Omukai08}
{Omukai}, K., {Schneider}, R., \& {Haiman}, Z. 2008, ArXiv 0804.3141

\bibitem[{{Omukai} {et~al.}(2005){Omukai}, {Tsuribe}, {Schneider}, \&
  {Ferrara}}]{Omukai05}
{Omukai}, K., {Tsuribe}, T., {Schneider}, R., \& {Ferrara}, A. 2005, \apj, 626,
  627

\bibitem[{{Penston}(1969)}]{Penston69}
{Penston}, M.~V. 1969, \mnras, 144, 425

\bibitem[{{Regan} \& {Haehnelt}(2009)}]{Regan09}
{Regan}, J.~A., \& {Haehnelt}, M.~G. 2009, \mnras, 396, 343

\bibitem[{{Schleicher} {et~al.}(2009){Schleicher}, {Galli}, {Glover},
  {Banerjee}, {Palla}, {Schneider}, \& {Klessen}}]{Schleicher09b}
{Schleicher}, D.~R.~G., {Galli}, D., {Glover}, S.~C.~O., {Banerjee}, R.,
  {Palla}, F., {Schneider}, R., \& {Klessen}, R.~S. 2009, \apj, 703, 1096

\bibitem[{{Schleicher} {et~al.}(2010){Schleicher}, {Spaans}, \&
  {Klessen}}]{SchleicherSpaans10}
{Schleicher}, D.~R.~G., {Spaans}, M., \& {Klessen}, R.~S. 2010, ArXiv e-prints
  1001.2118

\bibitem[{{Shang} {et~al.}(2009){Shang}, {Bryan}, \& {Haiman}}]{Shang09}
{Shang}, C., {Bryan}, G.~L., \& {Haiman}, Z. 2009, \mnras, 1840

\bibitem[{{Shapiro}(2005)}]{Shapiro05}
{Shapiro}, S.~L. 2005, \apj, 620, 59

\bibitem[{{Spaans} \& {Silk}(2006)}]{Spaans06}
{Spaans}, M., \& {Silk}, J. 2006, \apj, 652, 902

\bibitem[{{Wise} \& {Abel}(2007)}]{Wise07a}
{Wise}, J.~H., \& {Abel}, T. 2007, \apj, 665, 899

\bibitem[{{Wise} {et~al.}(2008){Wise}, {Turk}, \& {Abel}}]{Wise08a}
{Wise}, J.~H., {Turk}, M.~J., \& {Abel}, T. 2008, \apj, 682, 745

\bibitem[{{Yoshida} {et~al.}(2008){Yoshida}, {Omukai}, \&
  {Hernquist}}]{Yoshida08}
{Yoshida}, N., {Omukai}, K., \& {Hernquist}, L. 2008, Science, 321, 669

\end{thebibliography}
\end{document}